\pdfoutput=1

\documentclass[preprint,fleqn,5p,numbers,sort&compress]{elsarticle}

\usepackage{graphicx}
\usepackage{amssymb,amsmath}
\usepackage{natbib}
\usepackage{hyperref}
\hypersetup{pdftitle = The title of my PDF, pdfauthor = My name, pdfsubject= The subject, pdfkeywords = keyword1 keyword2 keyword3}
\hypersetup{colorlinks = true, linkcolor = blue, anchorcolor = red, citecolor = blue, filecolor = red, pagecolor = red, urlcolor = blue}

\journal{International Journal of Non-Linear Mechanics}

\begin{document}

\begin{frontmatter}

\title{On the Newton-Raphson basins of convergence of the out-of-plane equilibrium points in the Copenhagen problem with oblate primaries}

\author{Euaggelos E. Zotos\corref{cor1}}
\ead{evzotos@physics.auth.gr}

\cortext[cor1]{Corresponding author}

\address{Department of Physics, School of Science, Aristotle University of Thessaloniki, GR-541 24, Thessaloniki, Greece}

\begin{abstract}
The Copenhagen case of the circular restricted three-body problem with oblate primary bodies is numerically investigated by exploring the Newton-Raphson basins of convergence, related to the out-of-plane equilibrium points. The evolution of the position of the libration points is determined, as a function of the value of the oblateness coefficient. The attracting regions, on several types of two-dimensional planes, are revealed by using the multivariate Newton-Raphson iterative method. We perform a systematic and thorough investigation in an attempt to understand how the oblateness coefficient affects the geometry of the basins of convergence. The convergence regions are also related with the required number of iterations and also with the corresponding probability distributions. The degree of the fractality is also determined by calculating the fractal dimension and the basin entropy of the convergence planes.
\end{abstract}

\end{frontmatter}

\section{Introduction}
\label{intro}

The classical circular restricted three-body problem still remains, without any doubt, one of the most intriguing and open topics in celestial mechanics and dynamical astronomy. According to \cite{S67} the restricted three-body problem describes the motion of a third body, with an infinitesimal mass (thus acting as a test particle), inside the combined gravitational field of two primary bodies. This topic has numerous practical applications which expand from molecular physics, to chaos theory, planetary physics, as well as to galactic dynamics.

Over the last decades, the classical three-body problem has been substantially modified in an attempt to describe more realistically the nature of motion of massless test particles in the Solar System, by taking into consideration additional dynamical parameters of the system. In particular, the effective potential of the classical restricted three-body problem has been upgraded by including several types of additional forces.

The two primaries are spherical and homogeneous in the classical version of the restricted three-body problem. However, several celestial bodies in our Solar System (e.g., Saturn and Jupiter) have in fact an oblate shape \cite{BPC99}. In order to obtain a much more realistic description of the motion of the test particle in the vicinity of such oblate bodies the parameter of the oblateness has been introduced. The influence of the oblateness on the character of motion has been investigated in a series of papers (e.g., \cite{AEL12,BS12,KMP05,KDP06,KPP08,MPP96,MRVK00,PPK12,SSR75,SSR76,SSR79,SSR86,SL12,SL13,SRS88,SRS97,Z15a,Z15b}).

Another issue of great importance in dynamical systems is the so-called ``basins of convergence" associated to the equilibrium points. These convergence regions reveal how each point on a two-dimensional plane is attracted by the equilibrium points of the system, when an iterative method is used for numerically solving the system of the first order derivatives of the effective potential function. In the literature there is a plethora of numerical methods for numerically solving an equation with only one variable. For a system of equations, with two or more variables, on the other hand only a couple of methods exist. The most famous one is the classical Newton-Raphson method, while there is also the Broyden's method \cite{B65}, which however is in fact a quasi-Newton method. In numerous previous studies the Newton-Raphson iterative scheme has been used for determining the corresponding basins of convergence in several types of Hamiltonian systems (e.g., the Hill problem with oblateness and radiation pressure \cite{D10,Z17a}, the restricted three-body problem, where the primaries are magnetic dipoles \cite{KGK12}, the restricted three-body problem with oblateness and radiation pressure \cite{Z16}, the restricted four-body problem \cite{BP11,KK14,SAA17,SAP17}, the restricted five-body problem \cite{ZS18}, the ring problem of $N+1$ bodies \cite{CK07}, or even the pseudo-Newtonian restricted three-body problem \cite{Z17b}).

In dynamical system knowing the exact positions of the equilibrium points is an issue of paramount importance. Unfortunately, in many systems, such as those of the $N$-body problem (with $N \geq 3$), there are no explicit formulae for the positions of the libration points. Therefore, the locations of the equilibrium points can be obtained only by means of numerical methods. In other words, we need a multivariate iterative scheme for solving the system of the first order derivatives. It is well known that the results of any numerical method strongly depend on the initial conditions (staring points of the iterative procedure). Indeed, for some initial conditions the iterative formulae converge quickly, while for other starting points a considerable amount of iterations is required for reaching to a root (equilibrium point). Fast converging points usually belong to basins of convergence, while on the other hand slow converging points are located in fractal regions. On this basis, the knowledge of the basins of convergence of a dynamical system is very important because these basins reveal the optimal (regarding fast convergence) starting points for which the iterative formulae require the lowest amount of iterations, for leading to an equilibrium point. In addition, being aware of the fractal regions we know exactly which points should be avoided as initial conditions of the iterative formulae. At this point, it should be emphasized that the convergence properties of a dynamical system are directly linked to the chosen iterative formula. This implies that the basins of convergence will be different in case of another numerical method (e.g., Broyden's method).

It is well known that in the classical restricted three-body problem five coplanar equilibrium points exist \cite{S67}. In \cite{DM06} it was proved that in the case of oblate primary bodies there are four additional out-of-plane libration points. In the present study we will explore how the oblateness coefficient influences the position of these out-of-plane libration points as well as their corresponding basins of convergence. At this point it should be emphasized that there are no previous studies on the convergence areas of these out-of-plane points and therefore our analysis will shed some light, for the first time, on the dynamical properties of these equilibrium points.

The paper has the following structure: the most important properties of the dynamical system are presented in Section \ref{des}. The parametric evolution of the position of the out-of-plane equilibrium points is investigated in Section \ref{eqpts}. The following Section contains the main numerical results, regarding the structure of the Newton-Raphson basins of convergence, while in Section \ref{frac} we demonstrate how the oblateness coefficient affects the fractal dimension and the basin entropy. Our paper ends with Section \ref{conc}, where we emphasize the main conclusions of this work.

\section{Description of the Hamiltonian system}
\label{des}

The Hamiltonian system consists of two primary bodies, $P_1$ and $P_2$, which perform circular Keplerian orbits around their common mass center \cite{S67}. The third body moves under the combined gravitational attraction of the two primaries. Considering that the mass of the third body $m$ is considerable smaller, with respect to the masses of the two primary bodies $m_1$ and $m_2$, we may reasonably assume the motion of the primaries is not perturbed, in any way, by the test particle.

\begin{figure}
\centering
\resizebox{\hsize}{!}{\includegraphics{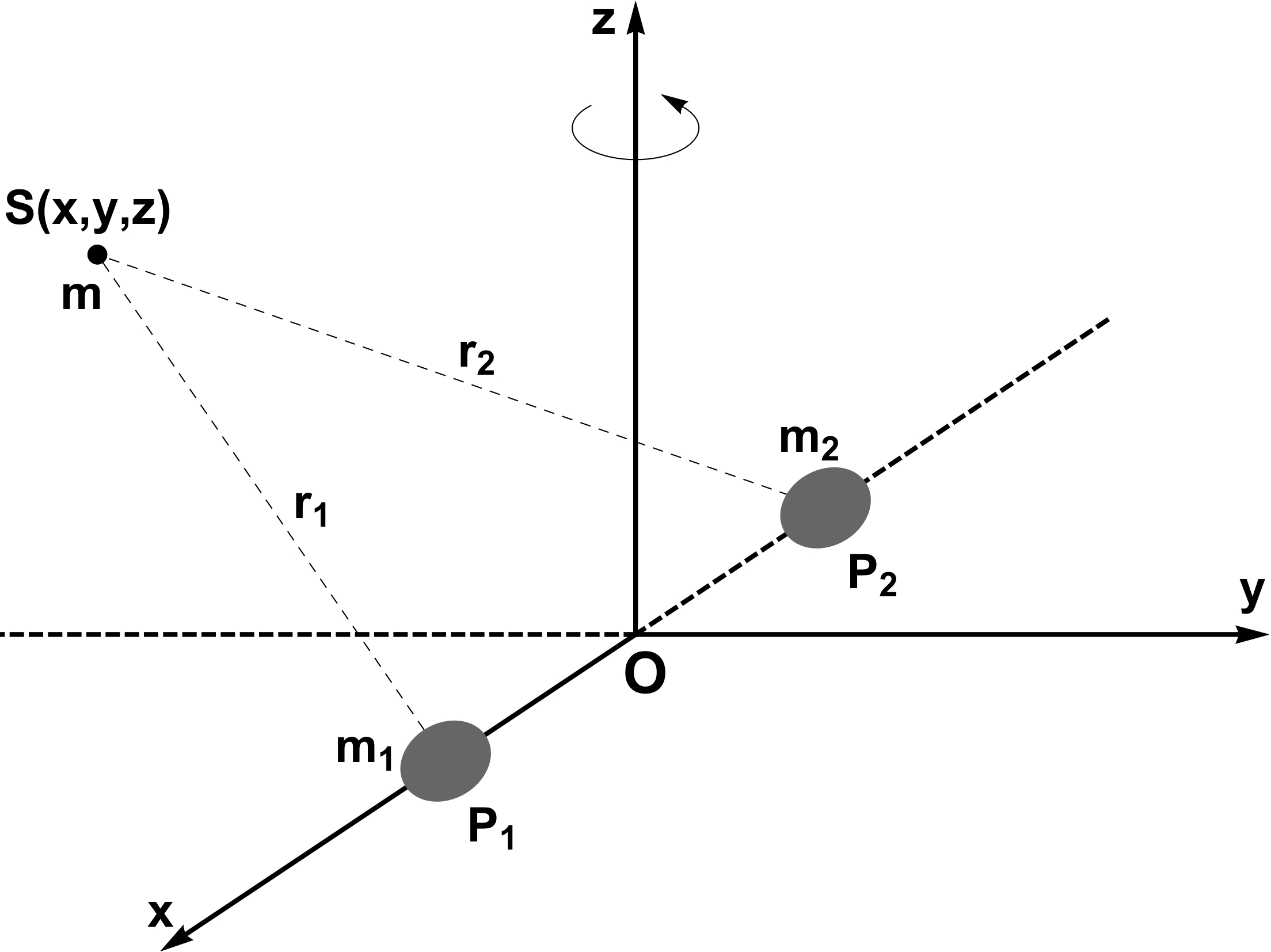}}
\caption{A schematic depicting the space configuration of the circular restricted three-body problem, when the primary bodies are oblate spheroids.}
\label{diag}
\end{figure}

In our system of units the gravitational constant $G$, the distance $R$ between the primaries and the sum of their masses are equal to unity. The dimensionless masses of the primary bodies are $m_1 = 1 - \mu$ and $m_2 = \mu$, where of course $\mu = m_2/(m_1 + m_2) \leq 1/2$ is the mass parameter. Furthermore, the centers of both primary bodies lie on the $x$-axis, at $(x_1, 0, 0)$ and $(x_2, 0, 0)$, where $x_1 = - \mu$ and $x_2 = 1 - \mu$. We consider a dimensionless, barycentric, rotating system of coordinates $Oxyz$, in which the $Ox$ axis always contains the two primary bodies, while the center of mass coincides with the origin $(0,0)$ (see Fig. \ref{diag}).

According to \cite{AS06,DM06,OV03,SSR75} the time-independent effective potential function of the restricted three-body problem with oblate primaries is
\begin{equation}
\Omega(x,y,z) = \sum_{n=1}^{2} \frac{m_i}{r_i}\left(1 + \frac{A_i}{2r_i^2} - \frac{3A_i z^2}{2r_i^4}\right) + \frac{n^2}{2} \left(x^2 + y^2 \right),
\label{pot}
\end{equation}
where
\begin{align}
r_1 &= \sqrt{\left(x - x_1 \right)^2 + y^2 + z^2}, \nonumber\\
r_2 &= \sqrt{\left(x - x_2 \right)^2 + y^2 + z^2},
\label{dist}
\end{align}
are the distances of the third body from the respective primaries. Moreover, $A_i$, $i = 1, 2$ are the oblateness coefficients, while $n$ is the mean motion which is given by
\begin{equation}
n = \sqrt{1 + \frac{3}{2}\left(A_1 + A_2 \right)}.
\label{vel}
\end{equation}
The exact derivation of equation (\ref{vel}), regarding the mean motion $n$ in the case of oblate primary bodies, is presented in the Appendix.

In this work, we consider only the case where the primaries are oblate spheroids $(A > 0)$, which means that the numerical values of the oblateness coefficients lie in the interval $[0,0.5]$.

The equations describing the motion of the test particle, in the corotating frame of reference, read
\begin{equation}
\ddot{x} - 2 n \dot{y} = \frac{\partial \Omega}{\partial x}, \ \ \
\ddot{y} + 2 n \dot{x} = \frac{\partial \Omega}{\partial y}, \ \ \
\ddot{z} = \frac{\partial \Omega}{\partial z}.
\label{eqmot}
\end{equation}

The Jacobi integral of motion is described by the Hamiltonian
\begin{equation}
J(x,y,z,\dot{x},\dot{y},\dot{z}) = 2\Omega(x,y,z) - \left(\dot{x}^2 + \dot{y}^2 + \dot{z}^2 \right) = C,
\label{ham}
\end{equation}
where $\dot{x}$, $\dot{y}$, and $\dot{z}$ are the velocities, while $C$ is the conserved value of the Hamiltonian.

\section{Out-of-plane equilibrium points}
\label{eqpts}

To what follows we will try to determine how the oblateness coefficient influences all the dynamical properties of the out-of-plane equilibrium points. In order to be absolutely sure that the changes on the properties are directly related to the oblateness we shall consider the Copenhagen case, where the two primary bodies have equal masses $(m_1 = m_2 = 1/2)$ and equal oblateness $A_1 = A_2 = A$.

The necessary and sufficient conditions, which must be fulfilled for the existence of equilibrium points, are
\begin{equation}
\dot{x} = \dot{y} = \dot{z} = \ddot{x} = \ddot{y} = \ddot{z} = 0.
\label{lps0}
\end{equation}
The corresponding coordinates $(x,y,z)$ of the libration points can be determined by solving numerically the system of the first order derivatives
\begin{equation}
\Omega_x(x,y,z) = 0, \ \ \ \Omega_y(x,y,z) = 0, \ \ \ \Omega_z(x,y,z) = 0,
\label{lps}
\end{equation}
where
\begin{align}
\Omega_x(x,y,z) &= \frac{\partial \Omega}{\partial x} = - \sum\limits_{i=1}^2 \frac{m_i \widetilde{x_i}}{r_i^3} \left(1 + \frac{3 A_i}{2 r_i^2} - \frac{15 A_i z^2}{2 r_i^4} \right) \nonumber\\
&+ n^2 x, \nonumber\\
\Omega_y(x,y,z) &= \frac{\partial \Omega}{\partial y} = - \sum\limits_{i=1}^2 \frac{m_i y}{r_i^3} \left(1 + \frac{3 A_i}{2 r_i^2} - \frac{15 A_i z^2}{2 r_i^4} \right) \nonumber\\
&+ n^2 y, \nonumber\\
\Omega_z(x,y,z) &= \frac{\partial \Omega}{\partial z} = - \sum\limits_{i=1}^2 \frac{m_i z}{r_i^3} \left(1 + \frac{9 A_i}{2 r_i^2} - \frac{15 A_i z^2}{2 r_i^4} \right),
\label{der1}
\end{align}
while $\widetilde{x_i} = x - x_i$, with $i = 1,2$.

In the classical restricted three-body problem (that is when $A_1 = A_2 = 0$) there are five equilibrium points, which are also known as Lagrange points. All five equilibrium points are coplanar and they are located on the configuration $(x,y)$ plane with $z = 0$. The central point $L_1$ is located between the two primaries, $L_2$ is located at the right side of primary $P_2$ (with $x > 0$), while $L_3$ is located at the left side of primary $P_1$ (with $x < 0$). In addition $L_4$ has $y > 0$, while the libration point $L_5$ has $y < 0$.

In \cite{DM06} it was shown that in the case with oblateness (that is when $A_1 \neq 0$ and $A_2 \neq 0$) additional
equilibrium points exist. More precisely, there are four additional libration points located on the $(x,z)$ plane (with $y =
0$), above and below the centers of the two oblate primaries.

\begin{figure}[!t]
\centering
\resizebox{\hsize}{!}{\includegraphics{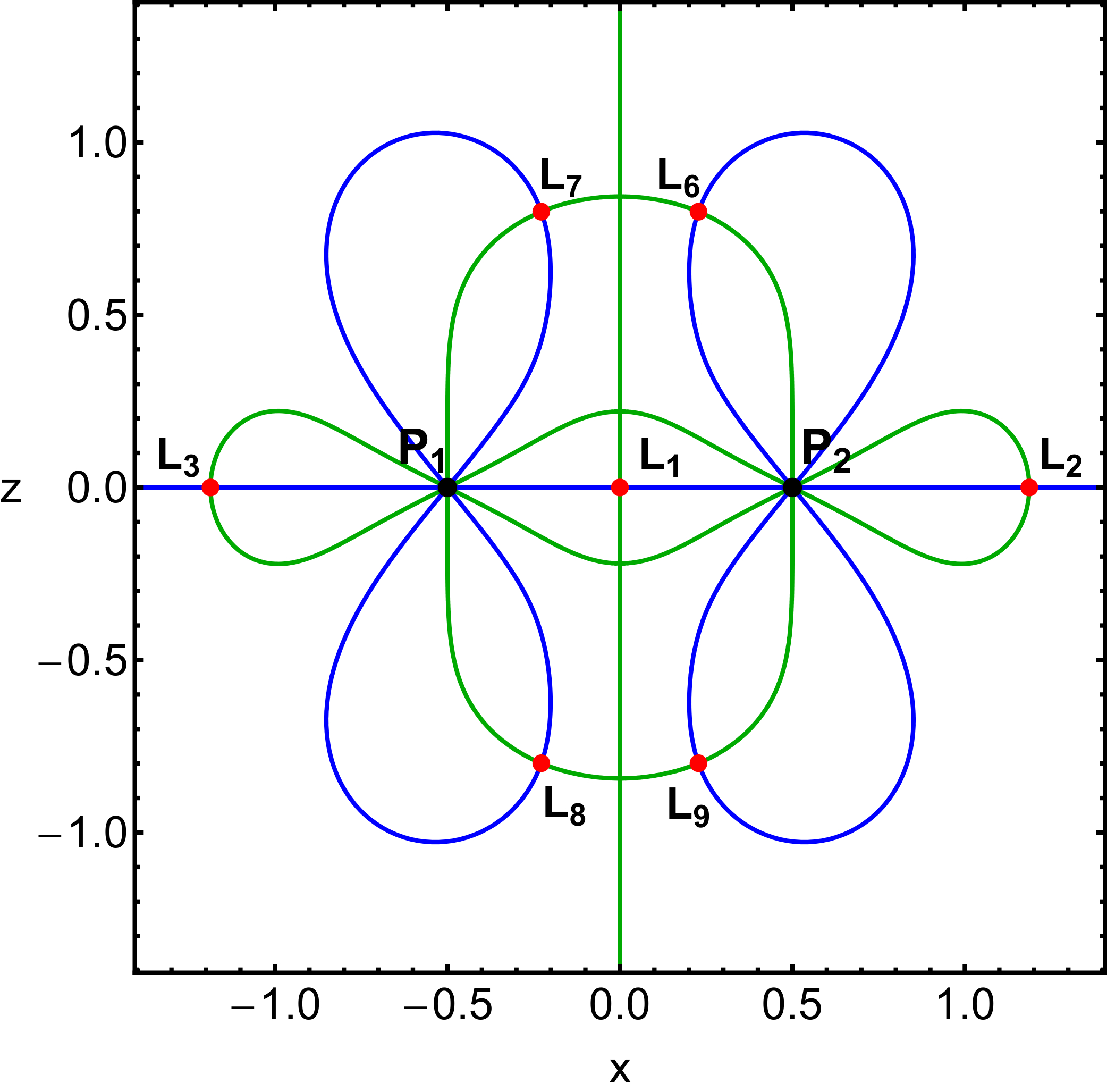}}
\caption{Positions (red dots) and numbering of the equilibrium points $(L_i, \ i=1,...,9)$ through the intersections of $\Omega_x = 0$ (green) and $\Omega_z = 0$ (blue), when $A = 0.5$. The black dots denote the two centers $(P_i, \ i=1,2)$ of the primaries. (Color figure online).}
\label{lgs}
\end{figure}

\begin{figure*}[!t]
\centering
\resizebox{0.75\hsize}{!}{\includegraphics{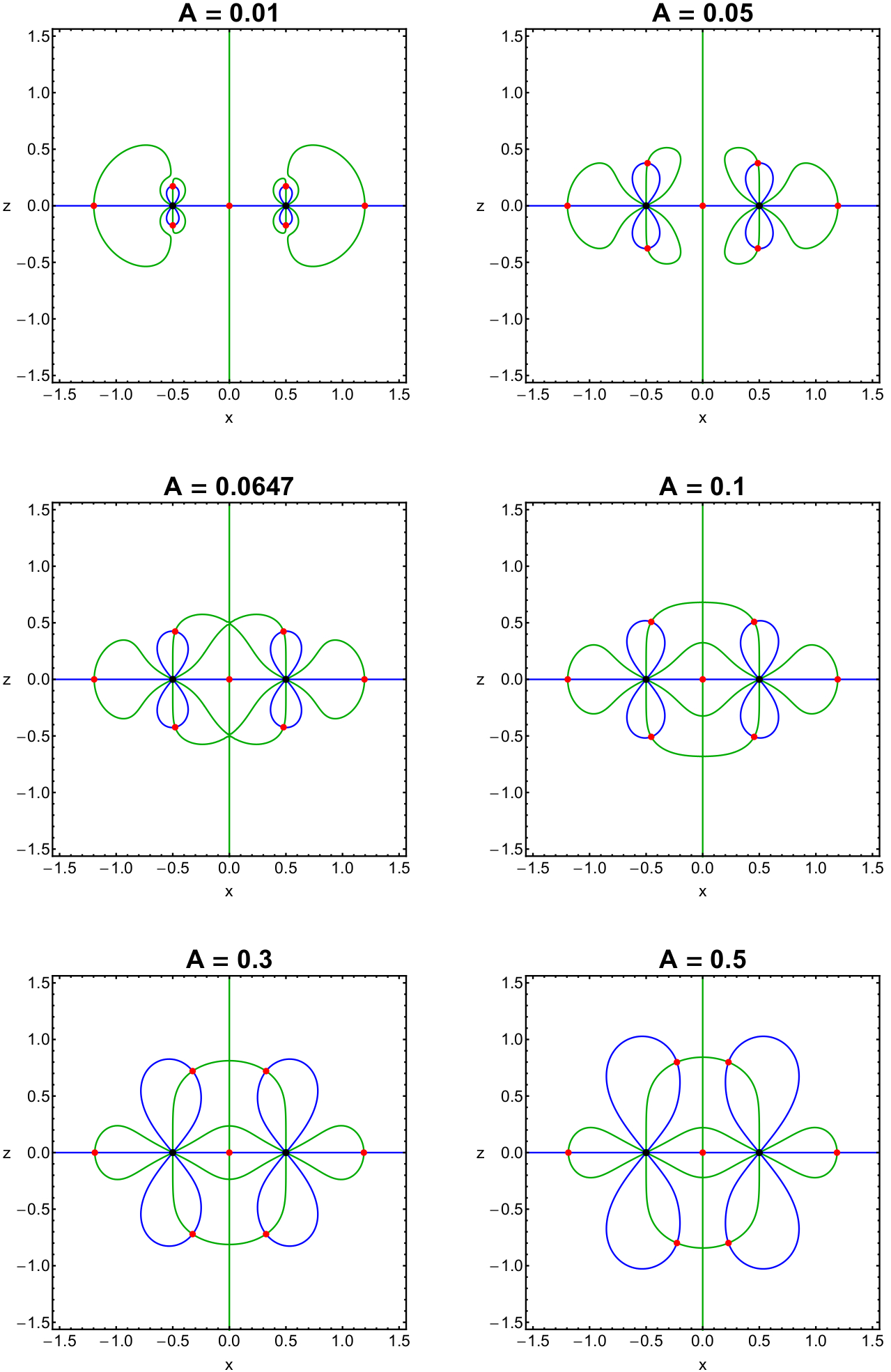}}
\caption{The variation of the positions of the equilibrium points (red dots) and the contours defined by the equations $\Omega_x = 0$ (green), $\Omega_z = 0$ (blue), as a function of the oblateness coefficient $A$. The black dots denote the two centers of the two primary oblate bodies. (Color figure online).}
\label{conts}
\end{figure*}

The intersections of the nonlinear equations $\Omega_x = 0$, and $\Omega_z = 0$ define the positions of the out-of-plane equilibrium points. Fig. \ref{lgs} illustrates how these equations pinpoint the location of the libration points, when $A = 0.5$. In the same diagram we explain the numbering, $L_i, \ i=6,...,9$, of all the out-of-plane equilibrium points. Note that the triangular points $L_4$ and $L_5$ are not visible on the $(x,z)$ plane, because for these points $x = z = 0$. In the following Fig. \ref{conts} we present how the positions of the equilibrium points, as well as the contours of the equations $\Omega_x = 0$, $\Omega_z = 0$ evolve, as a function of the value of the oblateness coefficient.

\begin{figure}[!t]
\centering
\resizebox{\hsize}{!}{\includegraphics{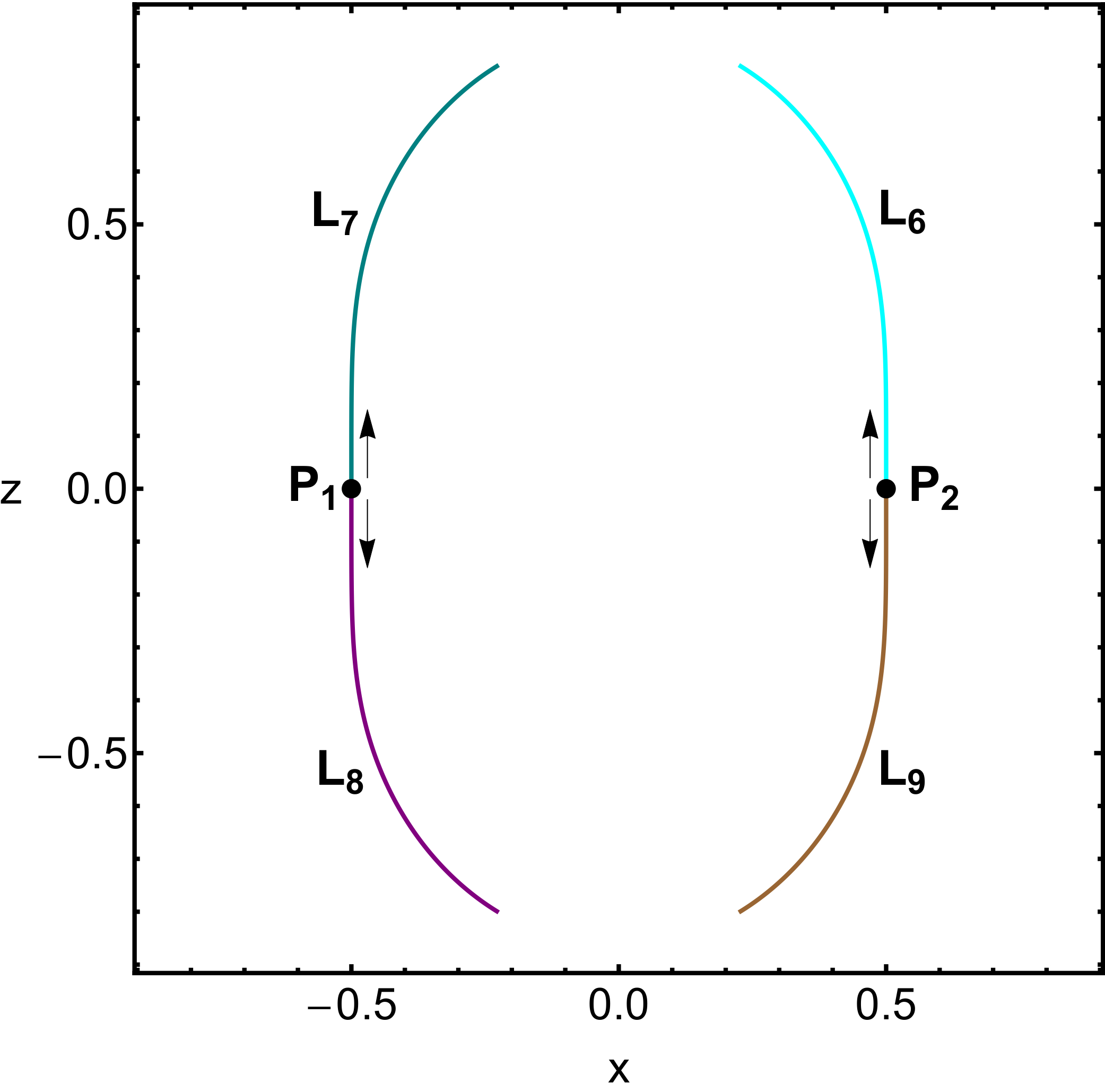}}
\caption{The parametric evolution of the positions of the out-of-plane equilibrium points, $L_i$, $i=6,...,9$, when $A \in (0, 0.5]$. The arrows indicate the movement direction of the equilibrium points as the value of the oblateness coefficient increases. The black dots pinpoint the fixed centers of the primaries. (Color figure online).}
\label{evol}
\end{figure}

The parametric evolution of the position of the out-of-plane equilibrium points, when $A \in (0,0.5]$ is shown in Fig. \ref{evol}. It is seen that as soon as $A > 0$ two pairs of out-of-plane equilibrium points appear just above the two centers $P_1$ and $P_2$. As the numerical value of the oblateness coefficient increases the out-of-plane equilibrium points start to move away from the centers. In particular, the absolute value of the $z$ coordinate increases, which means that they move away from the primary $(x,y)$ plane, while at the same time the absolute value of the $x$ coordinate decreases, which implies that they come close to the origin and the vertical $z$-axis. Here, we would like to note that the centers of the primary oblate bodies are completely unaffected by the shift of the oblateness coefficient.

\begin{figure}[!t]
\centering
\resizebox{\hsize}{!}{\includegraphics{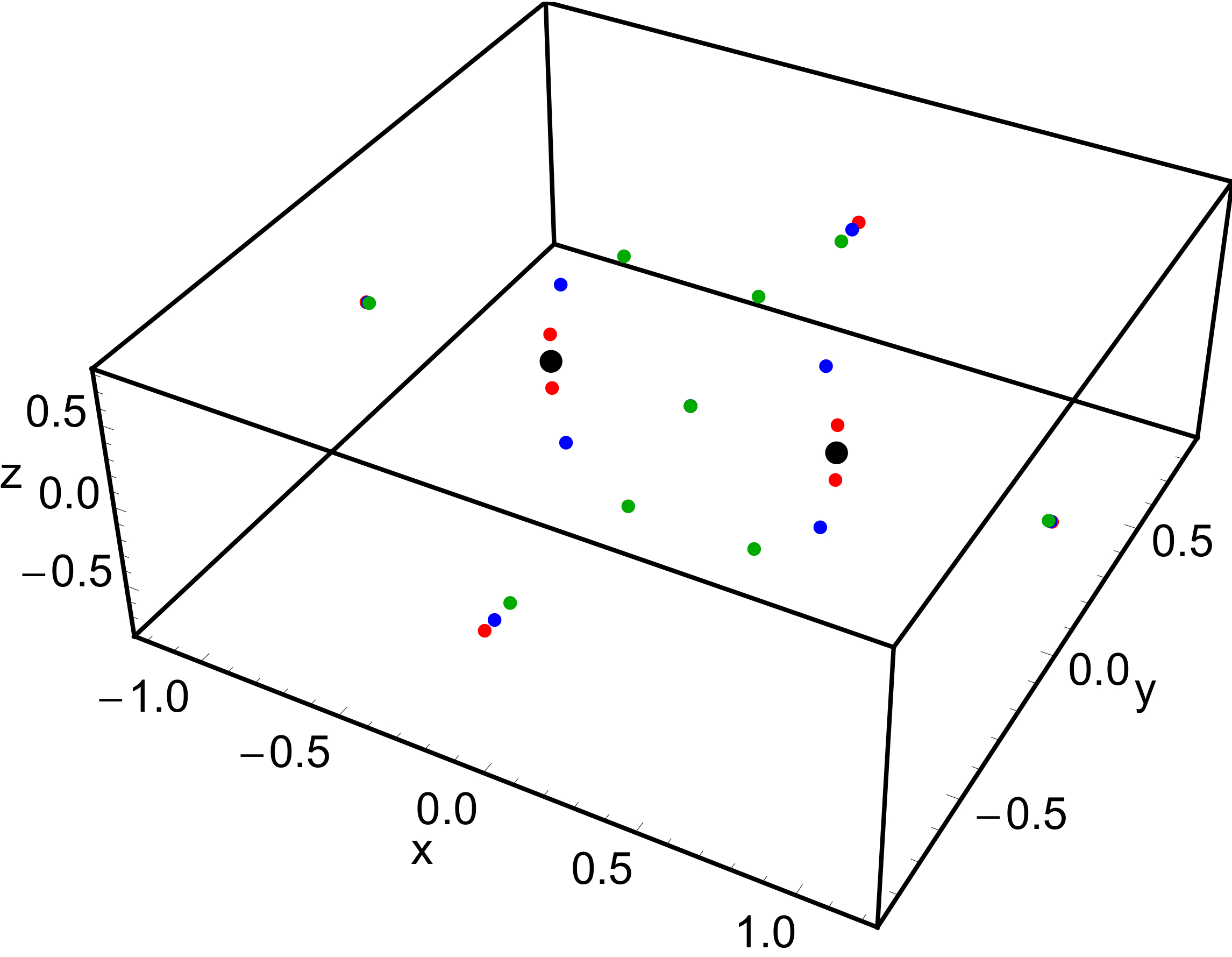}}
\caption{The spatial distribution of all the equilibrium points, when $A = 0.01$ (red), $A = 0.1$ (blue), and $A = 0.5$ (green). The fixed centers of the two primaries are indicated by black spheres. (Color figure online).}
\label{lgs3d}
\end{figure}

\begin{table}[!ht]
\begin{center}
   \caption{The coordinates for the set of the equilibrium points, presented in Fig. \ref{lgs3d}. Note that for all cases we have that $L_1: (0,0,0)$, $L_2: (x(L_2),0,0)$, $L_3: (-x(L_2),0,0)$, $L_4: (0,y(L_4),0)$, $L_5: (0,-y(L_4),0)$, $L_6: (x(L_6),0,z(L_6))$, $L_7: (-x(L_6),0,z(L_6))$, $L_8: (-x(L_6),0,-z(L_6))$, $L_9: (x(L_6),0,-z(L_6))$.}
   \label{tab1}
   \setlength{\tabcolsep}{5pt}
   \begin{tabular}{@{}lcccc}
      \hline
      $A$ & $x(L_2)$ & $y(L_4)$ & $x(L_6)$ & $z(L_6)$ \\
      \hline
      0.01 & 1.19759666 & 0.86044318 & 0.49969360 & 0.17276039 \\
      0.10 & 1.19284140 & 0.82325357 & 0.45475322 & 0.50805585 \\
      0.50 & 1.18683091 & 0.76349880 & 0.22789483 & 0.79916931 \\
      \hline
   \end{tabular}
\end{center}
\end{table}

\begin{table}[!ht]
\begin{center}
   \caption{The critical values of the Jacobi constant for the set of the equilibrium points, presented in Fig. \ref{lgs3d}. Note that for all cases we have that $C_2 = C_3$, $C_4 = C_5$, and $C_6 = C_7 = C_8 = C_9$.}
   \label{tab2}
   \setlength{\tabcolsep}{5pt}
   \begin{tabular}{@{}lcccc}
      \hline
      $A$ & $C_1$ & $C_2$ & $C_4$ & $C_6$ \\
      \hline
      0.01 & 4.08000000 & 3.51557655 & 2.78242742 & 5.09622526 \\
      0.10 & 4.80000000 & 4.04443099 & 3.06939775 & 2.42268824 \\
      0.50 & 8.00000000 & 6.39389258 & 4.30649015 & 1.41133794 \\
      \hline
   \end{tabular}
\end{center}
\end{table}

In Fig. \ref{lgs3d} we present the spatial distribution of all the equilibria for three values of the oblateness coefficient $A = {0.01, 0.1, 0.5}$. The values of the Jacobi constant at the equilibrium points are in fact critical values and they are denoted as $C_i$, with $i = 1, ..., 9$. In Tables \ref{tab1} and \ref{tab2} we provide the exact coordinates of the equilibrium points as well as the corresponding Jacobian constants, for the three cases, shown in Fig. \ref{lgs3d}.

Knowing the exact positions $(x_0,0,z_0)$ of the out-of-plane equilibrium points, we can easily determine their linear stability, through the nature of the six roots of the characteristic equation (see Eq. (12) in \cite{DM06}). Our computations indicate that the out-of-plane libration points are always unstable, when the oblateness coefficient $A$ varies in the interval $(0,0.5]$. Furthermore, additional numerical calculations suggest that the out-of-plane equilibrium points are universally linearly unstable for all the possible values of the mass parameter (when $\mu \in (0,0.5])$.

\section{The basins of convergence}
\label{bas}

There is no doubt that the most well-known numerical method for solving systems of nonlinear equations is the famous Newton-Raphson method. This method is applicable to systems of multivariate functions $f({\bf{x}}) = 0$ through the iterative scheme
\begin{equation}
{\bf{x}}_{n+1} = {\bf{x}}_{n} - J^{-1}f({\bf{x}}_{n}),
\label{sch}
\end{equation}
where $f({\bf{x_n}})$ denotes the system of equations, while $J^{-1}$ is the corresponding inverse Jacobian matrix, while in our case the system (\ref{lps}) contains three differential equations. It should be pointed out that the Newton-Raphson method can be also applied in systems with three equations. However the corresponding iterative scheme is very complicated. Therefore, in an attempt to make things simple we will exploit the fact that the out-of-plane equilibrium points lie on the $(x,z)$ plane, as in \cite{SAS18}. On this basis, we can use the bivariate Newton-Raphson scheme on the system
\begin{equation}
\Omega_x(x,0,z) = \Omega_z(x,0,z) = 0.
\label{sys1}
\end{equation}
Similarly, we can also use the bivariate Newton-Raphson scheme on the system
\begin{equation}
\Omega_x(0,y,z) = \Omega_y(0,y,z) = 0,
\label{sys2}
\end{equation}
for revealing the convergence properties of the $(y,z)$ plane.

For the $(x,z)$ plane the iterative formulae for each coordinate read
\begin{align}
x_{n+1} &= x_n - \left( \frac{\Omega_x \Omega_{zz} - \Omega_z \Omega_{xz}}{\Omega_{zz} \Omega_{xx} - \Omega^2_{xz}} \right)_{(x_n,z_n)}, \nonumber\\
z_{n+1} &= z_n + \left( \frac{\Omega_x \Omega_{zx} - \Omega_z \Omega_{xx}}{\Omega_{zz} \Omega_{xx} - \Omega^2_{xz}} \right)_{(x_n,z_n)},
\label{nrm1}
\end{align}
where $x_n$, $z_n$ are the values of the $x$ and $z$ coordinates at the $n$-th step of the iterative process. In the same vein, for the $(y,z)$ plane the corresponding iterative formulae are
\begin{align}
y_{n+1} &= y_n - \left( \frac{\Omega_y \Omega_{zz} - \Omega_z \Omega_{yz}}{\Omega_{zz} \Omega_{yy} - \Omega^2_{yz}} \right)_{(y_n,z_n)}, \nonumber\\
z_{n+1} &= z_n + \left( \frac{\Omega_y \Omega_{zy} - \Omega_z \Omega_{yy}}{\Omega_{zz} \Omega_{yy} - \Omega^2_{yz}} \right)_{(y_n,z_n)},
\label{nrm2}
\end{align}
where
\begin{align}
\Omega_{xx} &= \frac{\partial^2 \Omega}{\partial x^2} = - \sum\limits_{i=1}^2 \frac{m_i}{r_i^3} \Bigg( 1 + \frac{3\left(A_i - 2\widetilde{x_i}^2\right)}{2r_i^2} \nonumber\\
&-\frac{15 A_i \left(\widetilde{x_i}^2 + z^2 \right)}{2r_i^4} + \frac{105A_i \widetilde{x_i}^2 z^2}{2r_i^6} \Bigg) + n^2, \nonumber\\
\Omega_{xz} &= \frac{\partial^2 \Omega}{\partial x \partial z} = 3 \sum\limits_{i=1}^2 \frac{m_i \widetilde{x_i} z}{r_i^5} \left( 1 + \frac{15A_i}{2r_i^2} - \frac{35A_i z^2}{2r_i^4} \right), \nonumber\\
\Omega_{yy} &= \frac{\partial^2 \Omega}{\partial y^2} = - \sum\limits_{i=1}^2 \frac{m_i}{r_i^3} \Bigg( 1 + \frac{3\left(A_i - 2y^2\right)}{2r_i^2} \nonumber\\
&- \frac{15 A_i \left(y^2 + z^2 \right)}{2r_i^4} + \frac{105A_i y^2 z^2}{2r_i^6} \Bigg) + n^2, \nonumber\\
\Omega_{yz} &= \frac{\partial^2 \Omega}{\partial y \partial z} = 3 \sum\limits_{i=1}^2 \frac{m_i y z}{r_i^5} \left( 1 + \frac{15A_i}{2r_i^2} - \frac{35A_i z^2}{2r_i^4} \right), \nonumber\\
\Omega_{zx} &= \frac{\partial^2 \Omega}{\partial z \partial x} = \Omega_{xz} \nonumber\\
\Omega_{zy} &= \frac{\partial^2 \Omega}{\partial z \partial y} = \Omega_{yz} \nonumber\\
\Omega_{zz} &= \frac{\partial^2 \Omega}{\partial z^2} = - \sum\limits_{i=1}^2 \frac{m_i}{r_i^3} \Bigg( 1 + \frac{3\left(3A_i - 2z^2 \right)}{2r_i^2} \nonumber\\
&- \frac{45 A_i z^2}{r_i^4} + \frac{105 A_i z^4}{2 r_i^6} \Bigg).
\label{der2}
\end{align}

The philosophy behind the Newton-Raphson method is the following: An initial condition $(x_0,z_0)$ or $(y_0,z_0)$ activates the code, while the iterative procedure continues until an equilibrium point (attractor) is reached, with the desired predefined accuracy. If the particular initial condition leads to one of the libration points of the system it means that the numerical method converges for that particular initial condition. At this point, it should be emphasized that in general terms the method does not converge equally well for all the available initial conditions. The sets of the initial conditions which lead to the same attractor compose the so-called Newton-Raphson basins of attraction or basins of convergence or even attracting domains/regions.

At this point we must emphasize and clarify the following: the Newton-Raphson basins of convergence should not be mistaken with the basins of attraction which are present in dissipative systems. In dissipative systems we have the case of physical attractors. A physical attractor is a set of numerical values toward which a system tends to evolve, for a wide variety of initial conditions. On the other hand, for the case of an iterative scheme (e.g., the Newton-Raphson) we have the case of numerical attractors. A numerical attractor is a point (usually an equilibrium point) to which the iterative scheme leads for specific initial conditions. Obviously, a numerical attractor is not related, by any means, to a physical attractor, even though it behaves as such during the convergence process. For dissolving all confusion, we stress out that in this article we deal only with numerical attractors and their corresponding basins of convergence.

A double scan of the $(x,z)$ and $(y,z)$ planes is performed for revealing the structures of the basins of convergence. In particular, a dense uniform grid of $1024 \times 1024$ nodes is defined, in each type of plane, which shall be used as initial conditions of the iterative scheme. Evidently, the initial conditions of the centers of the primary bodies are of course excluded from all the grids because for these initial conditions the distances $r_i$, $i = 1,2$ to the respective primaries are equal to zero and therefore several terms, entering formulae (\ref{nrm1}) and (\ref{nrm2}), become singular. The number $N$ of the iterations, required for obtaining the desired accuracy, is also monitored during the classification of the nodes. For our computations, the maximum allowed number of iterations is $N_{\rm max} = 500$, while the iterations stop only when an attractor is reached, with accuracy of $10^{-15}$.

In the following subsections we will determine how the oblateness coefficient $A$ affects the structure of the Newton-Raphson basins of convergence in the circular restricted three-body problem with oblateness, by considering two cases regarding the type of the planes. For the classification of the nodes on each type of plane we will use color-coded diagrams (CCDs), in which each pixel is assigned a different color, according to the final state (attractor) of the corresponding initial condition.

\subsection{Results for the $(x,z)$ plane}
\label{ss1}

\begin{figure*}[!t]
\centering
\resizebox{0.9\hsize}{!}{\includegraphics{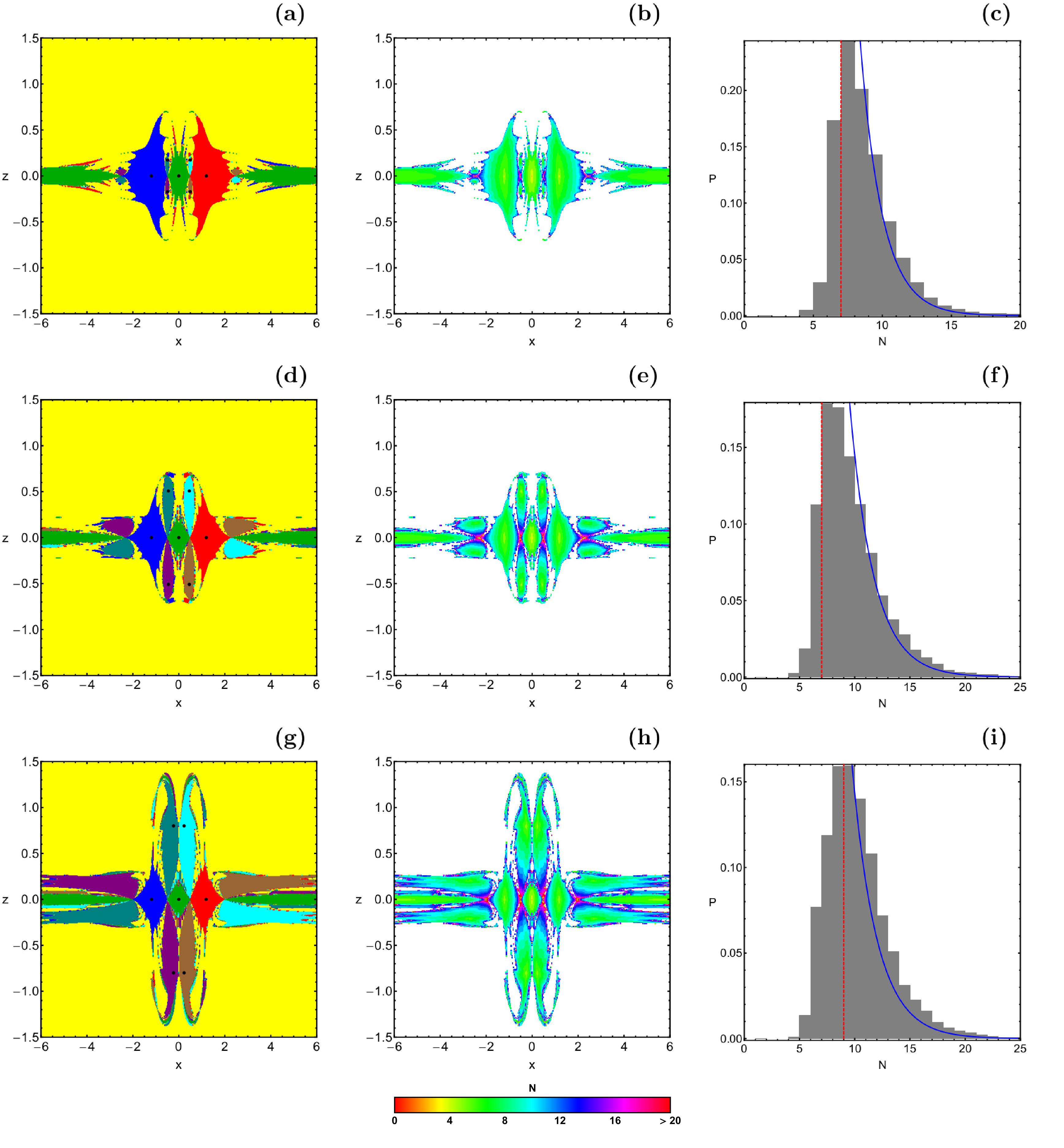}}
\caption{(First column): The Newton-Raphson basins of convergence on the $(x,z)$ plane. The color code, denoting the seven possible equilibrium points, is as follows: $L_1$ (green); $L_2$ (red); $L_3$ (blue); $L_6$ (cyan); $L_7$ (teal); $L_8$ (purple); $L_9$ (brown); tending to infinity (yellow); non-converging points (white). The positions of the seven libration points are indicated by black dots. (Second column): The distribution of the corresponding number $N$ of required iterations for obtaining the Newton-Raphson basins of convergence. The points tending to infinity as well as the non-converging points are shown in white. (Third column): The corresponding probability distribution of required iterations for obtaining the Newton-Raphson basins of convergence. The vertical dashed red line indicates, in each case, the most probable number $N^{*}$ of iterations. (First row): $A = 0.01$; (Second row): $A = 0.1$; (Third row): $A = 0.5$. (Color figure online).}
\label{m1}
\end{figure*}

\begin{figure*}[!t]
\centering
\resizebox{0.9\hsize}{!}{\includegraphics{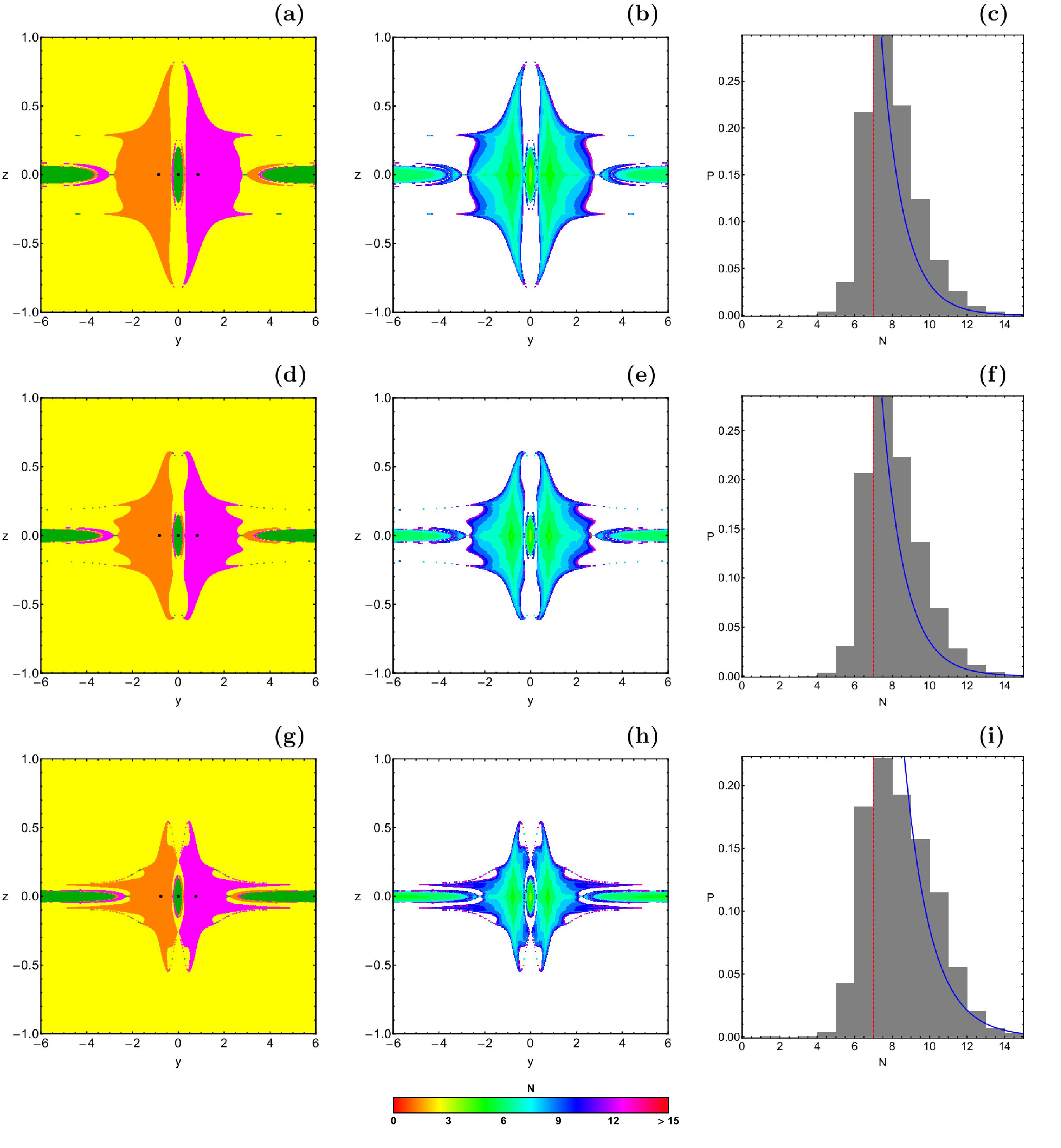}}
\caption{(First column): The Newton-Raphson basins of convergence on the $(y,z)$ plane. The color code, denoting the three possible equilibrium points, is as follows: $L_1$ (green); $L_2$ (red); $L_3$ (blue); tending to infinity (yellow); non-converging points (white). The positions of the three libration points are indicated by black dots. (Second column): The distribution of the corresponding number $N$ of required iterations for obtaining the Newton-Raphson basins of convergence. The points tending to infinity as well as the non-converging points are shown in white. (Third column): The corresponding probability distribution of required iterations for obtaining the Newton-Raphson basins of convergence. The vertical dashed red line indicates, in each case, the most probable number $N^{*}$ of iterations. (First row): $A = 0.01$; (Second row): $A = 0.1$; (Third row): $A = 0.5$. (Color figure online).}
\label{m2}
\end{figure*}

\begin{figure*}[!t]
\centering
\resizebox{\hsize}{!}{\includegraphics{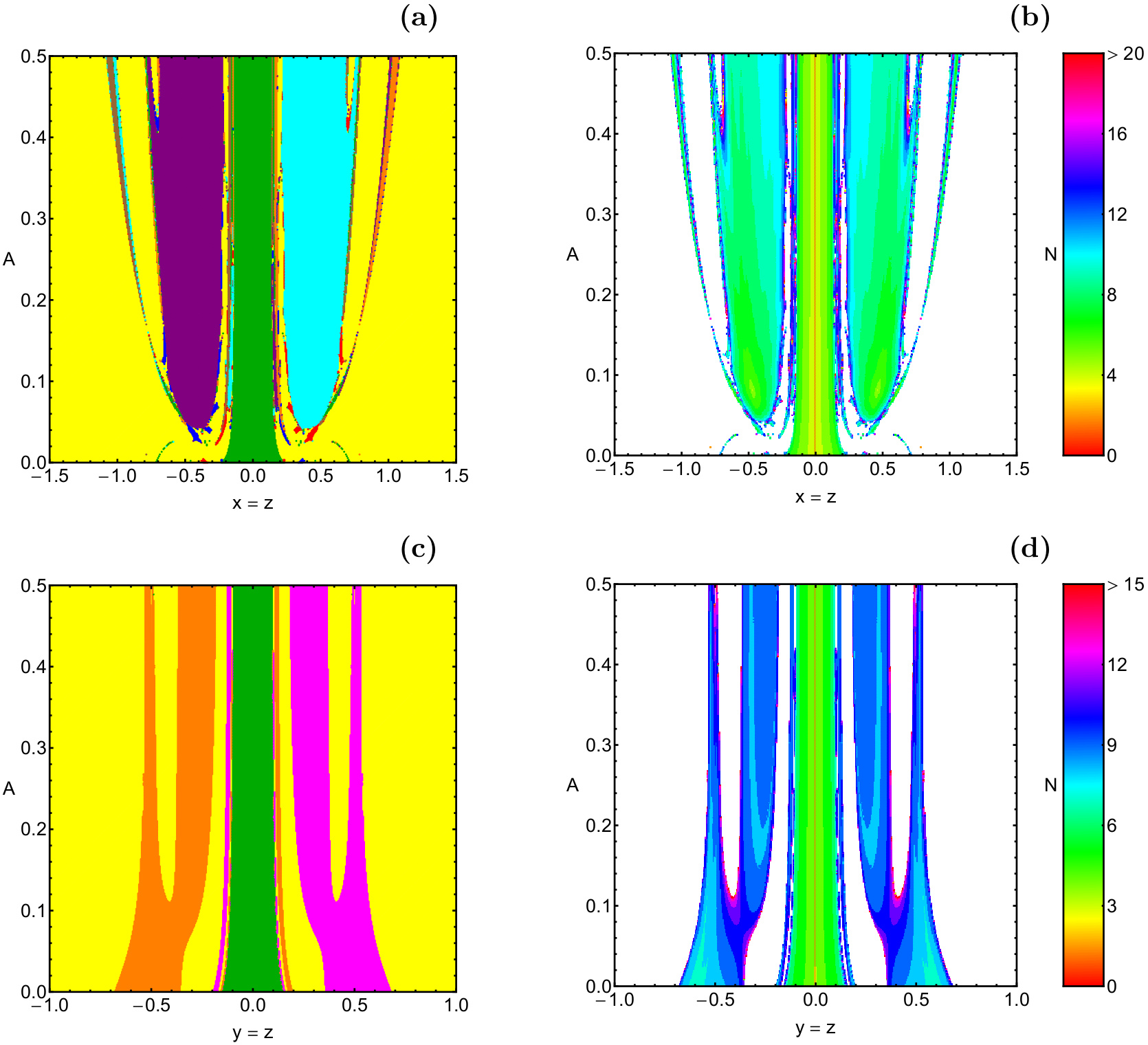}}
\caption{The Newton-Raphson basins of convergence on the (a-upper left): $(x = z,A)$ plane and (c-lower left): $(y = z,A)$ plane, when $A \in (0,0.5]$. The color code denoting the attractors is the same as in Figs. \ref{m1} and \ref{m2}. (Panels (b) and (d)): The distribution of the corresponding number $N$ of required iterations for obtaining the basins of convergence shown in panels (a) and (c), respectively. (Color figure online).}
\label{xyzA}
\end{figure*}

We begin our numerical investigation with the $(x,z)$ plane on which all the out-of-plane equilibrium points lie. The evolution of the geometry of the basins convergence, for three values of the oblateness coefficient, is illustrated in the first column of Fig. \ref{m1}. It is seen that in all cases the $(x,z)$ plane contains several well-defined basins of convergence which extend to infinity. On the other hand, the vast majority of the same plane is covered by initial conditions for which the Newton-Raphson scheme leads very quickly to extremely large numbers (yellow regions). For these initial conditions we may argue that we have strong numerical evidence that they asymptotically tend to infinity.

In the second column of Fig. \ref{m1} we present the corresponding number $N$ of iterations, using hue colors, while the corresponding probability distribution of the required iterations is given in the third column of the same figure. The definition of the probability $P$ is the following: if $N_0$ initial conditions $(x_0,z_0)$ converge, after $N$ iterations, to one of the equilibrium points then $P = N_0/N_t$, where $N_t$ is the total number of nodes in every CCD. Moreover, in all plots the tails of the histograms extend so as to cover 98\% of the corresponding distributions of iterations. The vertical, red, dashed line in the probability histograms denote the most probable number $N^{*}$ of iterations. The blue lines in the histograms of Fig. \ref{m1} indicate the best fit to the right-hand side $N > N^{*}$ of them (more details are given in subsection \ref{geno}).

With increasing value of the oblateness coefficient the most important changes, which occur on the $(x,z)$ plane, are the following:
\begin{itemize}
  \item The area of the basins of convergence, corresponding to collinear equilibrium points $L_1$, $L_2$, and $L_3$ is reduced, while at the same time the extent of the convergence regions, associated with the out-of-plane libration points $L_6$, $L_7$, $L_8$, and $L_9$, rapidly increases.
  \item The areas on the $(x,z)$ plane, for which the multivariate Newton-Raphson scheme requires a relatively high number of iterations $(N > 15)$, are reduced. Note that these areas are mainly located in the vicinity of the basin boundaries.
  \item The most probable number $N^{*}$ of iterations slightly increases from $N^{*} = 7$, when $A = 0.01$ to $N^{*} = 9$, when $A = 0.5$.
\end{itemize}

\subsection{Results for the $(y,z)$ plane}
\label{ss2}

On the $(y,z)$ plane only the three collinear equilibrium points $L_1$, $L_2$, and $L_3$, are visible, while on the other hand all the out-of-plane libration points are not present in this plane. However we feel that the information from this type of plane, along with the outcomes of the $(x,z)$ plane discussed earlier in the previous subsection, would help us to understand and obtain an initial draft idea, regarding the convergence properties of the entire three-dimensional $(x,y,z)$ space. In the first column of Fig. \ref{m2} we present the Newton-Raphson basins of convergence for three values of the oblateness coefficient.

As we proceed to higher values of the oblateness coefficient the main phenomena which take place, regarding the geometry of the convergence areas, are the following:
\begin{itemize}
  \item The extent of the convergence regions, corresponding to the central libration point $L_1$ decreases, while the area of the basins of convergence of the two triangular points constantly increases.
  \item In all studied cases, more than 98\% of the initial conditions converge, to one of three equilibrium points, within the first 15 iterations.
  \item The most probable number of iterations $N^{*}$ remains completely unperturbed at $N^{*} = 7$.
\end{itemize}

\subsection{A general overview}
\label{geno}

The color-coded convergence diagrams on the $(x,z)$ and $(y,z)$ planes, presented in Figs. \ref{m1} and \ref{m2}, provide sufficient information regarding the attracting domains, however for only a fixed value of the oblateness coefficient $A$. In order to overcome this handicap we can define a new type of distribution of initial conditions which will allow us to scan a continuous spectrum of $A$ values, rather than few discrete levels. The most interesting configuration is to set $x = z$ or $y = z$, while the value of the oblateness coefficient will vary in the interval $(0,0.5]$. This technique allows us to construct, once more, a two-dimensional plane in which the $x$, $y$ or the $z$ coordinate is the abscissa, while the value of $A$ is always the ordinate. Panels (a) and (c) of Fig. \ref{xyzA} show the basins of convergence on the $(x = z,A)$ plane, and $(y = z,A)$ plane, respectively, while in panels (b) and (d) of the same figure the distribution of the corresponding number $N$ of required iterations for obtaining the Newton-Raphson basins of convergence is shown.

Additional interesting information could be extracted from the probability distributions of iterations presented in the third row of the CCDs. In particular, it would be very interesting to try to obtain the best fit of the tails\footnote{By the term ``tails" of the distributions we refer to the right-hand side of the histograms, that is, for $N > N^{*}$.} of the distributions. For fitting the tails of the histograms, we used the Laplace distribution, which is the most natural choice, since this type of distribution is very common in systems displaying transient chaos (see e.g., \cite{ML01,SASL06,SS08}). Our calculations strongly indicate that in the vast majority of the cases the Laplace distribution is the best fit to our data.

The probability density function (PDF) of the Laplace distribution is given by
\begin{equation}
P(N | a,b) = \frac{1}{2b}
 \begin{cases}
      \exp\left(- \frac{a - N}{b} \right), & \text{if } N < a \\
      \exp\left(- \frac{N - a}{b} \right), & \text{if } N \geq a
 \end{cases},
\label{pdf}
\end{equation}
where $a$ is the location parameter, while $b > 0$, is the diversity. In our case we are interested only for the $x \geq a$ part of the distribution function.

\begin{figure*}[!t]
\centering
\resizebox{\hsize}{!}{\includegraphics{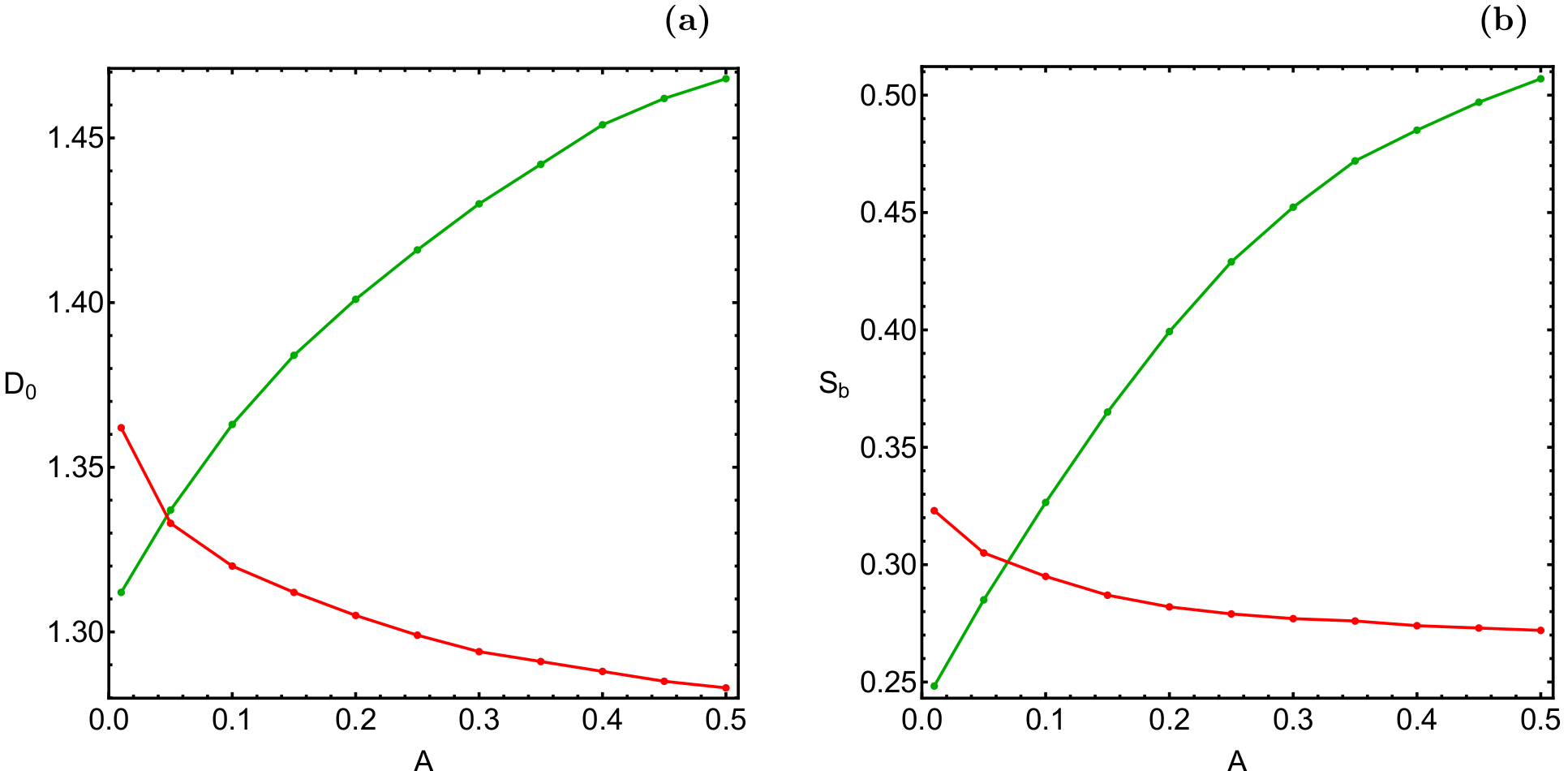}}
\caption{Evolution of the (a-left): fractal dimension $D_0$ and (b-right): basin entropy $S_b$, of the $(x,z)$ plane (green) and $(y,z)$ plane (red), as a function of the oblateness coefficient $A$.}
\label{fract}
\end{figure*}

In Table \ref{t1} we present the values of the location parameter $a$ and the diversity $b$, as they have been obtained through the best fit, for all cases discussed in the previous subsections. One may observe that for most of the cases the location parameter $a$ is very close to the most probable number $N^{*}$ of iterations, while in some cases these two quantities coincide.

\begin{table}[!ht]
\begin{center}
   \caption{The values of the location parameter $a$ and the diversity $b$, related to the most probable number $N^{*}$ of iterations, for all the studied cases shown earlier in the CCDs.}
   \label{t1}
   \setlength{\tabcolsep}{15pt}
   \begin{tabular}{@{}lcccc}
      \hline
      Figure & $\epsilon$ & $N^{*}$ & $a$ & $b$ \\
      \hline
      \ref{m1}c & 0.01 & 7 & $N^{*} + 1$ & 1.61 \\
      \ref{m1}f &  0.1 & 7 & $N^{*} + 2$ & 2.19 \\
      \ref{m1}i &  0.5 & 9 & $N^{*}$     & 2.23 \\
      \hline
      \ref{m2}c & 0.01 & 7 & $N^{*}$     & 1.18 \\
      \ref{m2}f &  0.1 & 7 & $N^{*}$     & 1.22 \\
      \ref{m2}i &  0.5 & 7 & $N^{*} + 1$ & 1.42 \\
      \hline
   \end{tabular}
\end{center}
\end{table}

\section{Parametric evolution of the fractality of the convergence planes}
\label{frac}

In the CCDs of the previous section we observed the presence of highly fractal regions, mainly located near the vicinity of the basin boundaries. It is known that the final state (equilibrium point) of initial conditions inside these fractal areas is highly sensitive. Specifically, even the slightest change of the initial conditions automatically leads to a completely different libration point, which is a classical indication of chaos. Therefore, for the initial conditions located in the basin boundaries it is almost impossible to predict their final states. By using the term fractal we simply imply that the particular areas display a fractal-like geometry, however without computing any quantitative parameter, such as the fractal dimension (e.g., \cite{AVS01,AVS09}).

So far we discussed the fractality of the convergence diagrams using only qualitative arguments. However it would be very informative if we could have quantitative results regarding the evolution of the fractality. In order to measure the degree of fractality we have computed the uncertainty dimension \cite{O93} for different values of the oblateness coefficient $A$, thus following the computational method introduced in \cite{AVS01}. Obviously, the degree of fractality is completely independent of the initial conditions we used to compute it. The evolution of the uncertainty dimension $D_0$ for both the $(x,z)$ and $(y,z)$ planes, as a function of the oblateness coefficient $A$, is shown in panel (a) of Fig. \ref{fract}. The computations of the uncertainty dimension were performed for two-dimensional grids of initial conditions and for that reason $D_0 \in (1,2)$, where $D_0 = 2$ means total fractality, while $D_0 = 1$ implies zero fractality. It is seen that, with increasing value of the oblateness coefficient, $D_0$ increases on the $(x,z)$ plane, while on the other hand it gradually decreases on the $(y,z)$ plane.

Very recently, in \cite{DWGGS16}, a new quantitative tool was introduced, for measuring the degree of the basin fractality. This new dynamical quantity is called ``basin entropy" and it measures the degree of fractality (or unpredictability) of the basins, by examining their topological properties.

The basin entropy works according to the following numerical algorithm. If there are $N(A)$ attractors (equilibrium points or roots) in a certain rectangular region $R = [-x_L,x_L] \times [-y_L,y_L]$ of the convergence plane (for the $(x,z)$ plane $x_L = 6$ and $y_L = 1.5$, while for the $(y,z)$ plane $x_L = 6$ and $y_L = 1$), then we subdivide $R$ into a grid of $N$ square boxes, where each cell of the gird may contain between 1 and $N(A)$ attractors. Then the probability that inside the cell $i$ the corresponding attractor is $j$ is denoted by $P_{i,j}$. Taking into account that inside each cell the initial conditions are completely independent, the Gibbs entropy, of every cell $i$ reads
\begin{equation}
S_{i} = \sum_{j=1}^{m_{i}}P_{i,j}\log_{10}\left(\frac{1}{P_{i,j}}\right),
\end{equation}
where $m_{i} \in [1,N_{A}]$ is the number of the attractors inside the box $i$.

The total entropy of the entire region $R$, on the configuration plane, can easily be calculated by adding the entropies of the $N$ cells of the grid as $S = \sum_{i=1}^{N} S_{i}$. Therefore, the total entropy, corresponding to the total number of cells $N$ is called basin entropy and it is given by
\begin{equation}
S_{b} = \frac{1}{N}\sum_{i=1}^{N}\sum_{j=1}^{m_{i}}P_{i,j}\log_{10}\left(\frac{1}{P_{i,j}}\right).
\end{equation}

Following the above-mentioned algorithm and also using the value $\varepsilon = 0.005$, suggested in \cite{DWGGS16}, we calculated the numerical value of the basin entropy $S_b$ of the $(x,z)$ and $(y,z)$ planes, when the oblateness coefficient lies in the interval $A \in (0,0.5]$. At this point, it should be clarified that in the case where non-converging points, or points that tend asymptotically to infinity are present, we count them as additional basins which coexist with the other basins, corresponding to the equilibrium points. In panel (b) of Fig. \ref{fract} we present the evolution of the basin entropy as a function of $A$. At this point, it should be noted that for creating this diagram we used numerical results not only for the three cases, presented earlier in Figs. \ref{m1} and \ref{m2}, but also from additional values of $A$.

It is observed in Fig. \ref{fract}(a-b) that both the uncertainty dimension and the basin entropy increase on the $(x,z)$ plane, while they decrease on the $(y,z)$ plane, with increasing value of the oblateness coefficient. This behavior can be explained by looking the corresponding convergence diagrams, given in Figs. \ref{m1} and \ref{m2}. More precisely it is seen that the area of the basins of convergence on the $(x,z)$ plane increases, while the area of the convergence regions on the $(y,z)$ plane decreases, as we proceed to higher values of $A$. Therefore, the portion of the fractal regions on the $(x,z)$ and $(y,z)$ planes increases and decreases, respectively. This directly implies that the degree of fractality (expressed through the uncertainty dimension and the basin entropy) displays a different parametric evolution on both types of planes.

Looking both panels of Fig. \ref{fract} we encounter a very interesting phenomenon. We refer of course to the very similar parametric evolution of the fractal dimension $D_0$ as well as the basin entropy $S_b$. It should be noted, that this is the first time that these two dynamical quantities are compared, using numerical results of the same system. We assume that the impressive similarity of their parametric evolution reflects the fact that both these dynamical quantities provide, in a different way, the same information, regarding the degree of fractality of a two-dimensional plane.

\section{Concluding remarks}
\label{conc}

We numerically explored the basins of convergence in the circular restricted three-body problem with oblate primary bodies. More precisely, we demonstrated how the oblateness coefficient $A$ influences the position of the out-of-plane equilibrium points. The multivariate Newton-Raphson iterative scheme was used for revealing the corresponding basins of convergence on the $(x,z)$ and $(y,z)$ planes. These convergence domains play a significant role, since they explain how each point is numerically attracted by the equilibrium points of the system. We managed to monitor how the Newton-Raphson basins of convergence evolve as a function of the oblateness coefficient. Another important aspect of this work was the relation between the basins of convergence and the corresponding number of required iterations and the respective probability distributions.

This is the first time that the Newton-Raphson basins of convergence, corresponding to the out-of-plane equilibrium points, are revealed as well as numerically investigated in such a systematic and thorough manner. On this basis, the presented results are novel and this is exactly the main contribution of our work. It should be noted that all the numerical results of this work (basins of convergence, degree of fractality, etc) refer to the specific numerical method (Newton-Raphson).

The following list contains the most important conclusions of our numerical analysis.
\begin{enumerate}
  \item It was found that all the basins of convergence, corresponding to all equilibrium points, have infinite area, regardless the value of the oblateness coefficient.
  \item Our numerical analysis indicates that the vast majority of the $(x,z)$ and $(y,z)$ planes is covered by initial conditions which do not converge to any of the libration points. Furthermore, additional computations revealed that for all these initial conditions the multivariate Newton-Raphson iterator lead very fast to extremely large numbers, which implies that these initial conditions tend asymptotically to infinity.
  \item It should be emphasized that our classification of the initial conditions on the two-dimensional planes did not report any non-converging nodes (initial conditions which do not converge after 500 iterations) or false-converging nodes to final states different, with respect to the equilibrium points of the system.
  \item In general terms, the Newton-Raphson method was found to converge very fast $(0 \leq N < 5)$ for initial conditions close to the roots, fast $(5 \leq N < 10)$ and slow $(10 \leq N < 15)$ for initial conditions that complement the central regions of the very fast convergence, and very slow $(N \geq 15)$ for initial conditions of dispersed points lying either in the vicinity of the basin boundaries, or between the dense regions of the equilibrium points.
  \item It was observed that with increasing value of the oblateness coefficient both the fractal dimension and the basin entropy increase on the $(x,z)$ plane, while they both decrease on the $(y,z)$ plane.
\end{enumerate}

A double precision numerical code, written in standard \verb!FORTRAN 77! \cite{PTVF92}, was used for the classification of the initial conditions into the different types of basins. In addition, for all the graphical illustration of the paper we used the latest version 11.2 of Mathematica$^{\circledR}$ \cite{W03}. Using an Intel$^{\circledR}$ Quad-Core\textsuperscript{TM} i7 2.4 GHz PC the required CPU time, for the classification of each set of initial conditions, was about 5 minutes.

\section{Future work}
\label{fut}

In the present paper we use the bivariate Newton-Raphson iterative scheme for revealing the corresponding basins of convergence on the two-dimensional $(x,z)$ and $(y,z)$ planes. However it is in our future plans to explore the convergence properties of the entire three-dimensional $(x,y,z)$ space, by numerically solving the system of the three equations (\ref{lps}). Currently, the development of a numerical code for solving simultaneously all three equations (\ref{lps}) is underway and we hope that very soon we will be able to demystify the secrets of the $(x,y,z)$ space.

The current results of the $(x,z)$ and $(y,z)$ planes provide sufficient information, regarding the convergence properties of the system. At this point it should be noted that even in the case where we will be able to numerically solve simultaneously all three equations the use of two-dimensional planes will be again the only feasible choice of visualizing the basins of convergence. This is true if we take into account that for a solid three-dimensional grid of initial conditions (inside the $(x,y,z)$ space) only its outer shell is visible. Therefore, the best approach, in order to visualize the inner structures of the basins structures, will be to use tomographic slices on several two-dimensional planes.

\section*{Compliance with Ethical Standards}

\begin{itemize}
  \item Funding: The author states that he has not received any research grants.
  \item Conflict of interest: The author declares that he has no conflict of interest.
\end{itemize}

\section*{Acknowledgments}

The author would like to express his warmest thanks to the two anonymous reviewers for the careful reading of the manuscript and for all the apt suggestions and comments which allowed us to improve both the quality and the clarity of the paper.

\begin{appendix}

\section{Derivation of the mean motion $n$}
\label{appex}

Let the distances of the primary bodies $P_1$ and $P_2$ from the origin $O$ be $a$ and $b$ respectively. Since $P_1$ and $P_2$ are moving in circular orbits the gravitational forces $F_{gi}$ are equal to the corresponding centrifugal forces $F_{ci}$, with $i = 1, 2$ acting on the two oblate primaries. In particular we have that $F_{c1} = F_{g1}$ and $F_{c2} = F_{g2}$. Adding these two equation and after trivial computations we obtain
\begin{align}
n^2 &= \frac{G \left(m_1 + m_2\right)}{\left(a + b\right)^3} + \frac{3 G \left(m_1 + m_2\right)}{2 m_1 \left(a + b\right)^5} S \nonumber\\
&+ \frac{3 G \left(m_1 + m_2\right)}{2 m_2 \left(a + b\right)^5} S',
\end{align}
where $S = \left(I_1 + I_2 + I_3 - 3 I\right)$ and $S' = \left(I'_1 + I'_2 + I'_3 - 3 I'\right)$. Here $I$ is the moment of inertia, of the body $P_1$, through the line joining the centre of mass of $P_1$ and $P_2$. Similarly, $I'$ is the moment of inertia, of the body $P_2$, through the line joining the centre of mass of $P_2$ and $P_1$. In the same vein, $I_i$ and $I'_i$, with $i = 1, ..., 3$ are the principle moments of inertia of the oblate bodies $P_1$ and $P_2$, respectively through their centers of mass.

In the dimensionless variables, where $G = m_1 + m_2 = a + b = 1$, we have
\begin{equation}
n^2 = 1 + \frac{3}{2 m_1} S + \frac{3}{2 m_2} S'.
\label{mm}
\end{equation}

It is known that
\begin{align}
I_1 &= I_2 = m_1\left(\frac{R_{1e}^2 + R_{1p}^2}{5R^2}\right), \nonumber\\
I_3 &= \frac{2 m_1 R_{1e}^2}{5R^2}, \nonumber\\
I &= I_1,
\end{align}
where $R_e$ and $R_p$ are the equatorial and the polar radius, respectively of the oblate primary $P_1$, while $R$ is the distance between the centers of the two primaries. Similar equations apply for the oblate body $P_2$.

Substituting the above-mentioned formulae to equation (\ref{mm}) we get
\begin{equation}
n^2 = 1 + \frac{3}{2}\left(\frac{R_{1e}^2 - R_{1p}^2}{5R^2}\right) + \frac{3}{2}\left(\frac{R_{2e}^2 - R_{2p}^2}{5R^2}\right).
\end{equation}
Therefore we obtain that
\begin{equation}
n = \sqrt{1 + \frac{3}{2}\left(A_1 + A_2\right)},
\label{mmf}
\end{equation}
where
\begin{align}
A_1 &= \frac{R_{1e}^2 - R_{1p}^2}{5R^2}, \nonumber\\
A_2 &= \frac{R_{2e}^2 - R_{2p}^2}{5R^2},
\end{align}
are the definitions of the oblateness coefficients of the primary bodies.

\end{appendix}

\end{document}